\documentclass[
aps,
prd,
noeprint,
longbibliography,
onecolumn,
superscriptaddress,
amsmath,amssymb,
floatfix
]{revtex4-2}
\usepackage[caption=false]{subfig} 
\usepackage{ragged2e} 
\DeclareCaptionJustification{justified}{\justifying}

\usepackage{amssymb,amsmath}
\usepackage{graphicx}
\usepackage{amsmath}		
\usepackage[margin=1.0in]{geometry}
\usepackage{setspace}
\usepackage{color}
\usepackage{fancyhdr}
\usepackage{collcell}
\usepackage{datatool}
\usepackage{latexsym}
\usepackage{amssymb}
\usepackage{epsfig,amsmath,graphics}
\usepackage{epstopdf}
\usepackage{verbatim}
\usepackage{wasysym}
\usepackage{feynmp-auto}
\usepackage{xcolor}
\usepackage{enumitem}
\usepackage[utf8]{inputenc}
\usepackage{slashed}

\usepackage{hyperref}
\hypersetup{
    colorlinks=true,
    linkcolor=blue,
    filecolor=magenta,      
    urlcolor=cyan,
    pdftitle={Overleaf Example},
    pdfpagemode=FullScreen,
    }

\usepackage{empheq}

    \newlength\fsep
    \newsavebox\widebox

\usepackage[skip=0pt]{caption}

\begin{document}
\title{Search for monopole-dipole interactions with atom interferometry}







\author{Mahiro Abe}
\affiliation{Department of Physics, Stanford University, Stanford, CA  
94305, USA}

\author{Jason M. Hogan}
\affiliation{Department of Physics, Stanford University, Stanford, CA  
94305, USA}

\author{David E.~Kaplan}
\affiliation{Department of Physics \& Astronomy, The Johns Hopkins University, Baltimore, MD  21218, USA}

\author{Chris Overstreet}
\affiliation{Department of Physics \& Astronomy, The Johns Hopkins University, Baltimore, MD  21218, USA}

\author{\mbox{Surjeet Rajendran}}
\affiliation{Department of Physics \& Astronomy, The Johns Hopkins University, Baltimore, MD  21218, USA}

\begin{abstract}
Light, weakly coupled bosonic particles such as axions can mediate long range monopole-dipole interactions between matter and spins. We propose a new experimental method using atom interferometry to detect such a force on a freely falling atom exerted by the spin of electrons. The intrinsic advantages of atom interferometry, such as the freely falling nature of the atom and the well-defined response of the atom to external magnetic fields, should enable the proposed method to overcome systematic effects induced by vibrations, magnetic fields, and gravity. This approach is most suited to probe forces with a range $\gtrsim$~10~cm. With current technology, our proposed setup could potentially extend probes of such forces by an order of magnitude beyond present laboratory limits. 
\end{abstract}

\maketitle

\section{Introduction}
\label{sec:intro}

The existence of light, weakly coupled bosonic particles is one of the generic features of attempts to solve outstanding problems of particle physics such as the hierarchy~\cite{Graham:2015cka}, strong CP~\cite{Peccei:1977hh, Weinberg:1977ma, Wilczek:1977pj} and flavor problems~\cite{Wilczek:1982rv}. The detection of such a boson could thus provide a unique window into high energy physics. This exciting possibility has motivated a significant experimental program aimed at detecting these bosons, either by sourcing the particle in the laboratory~\cite{Moody:1984ba} or by detecting a cosmic abundance of such particles~\cite{Preskill:1982cy, Dine:1982ah, Dine:1981rt}. Experiments that source and detect the particle in controlled laboratory conditions are particularly important because they are not reliant on unknown cosmic history to produce the particle and are thus a robust probe of its existence. There are two broad classes of such experiments. In the first, the new particles are produced on shell; this includes ``light shining through a wall,''~\cite{Ejlli:2020yhk} beam dump~\cite{Bjorken:2009mm}, and nuclear decay experiments~\cite{Benato:2018ijc, Dent:2021jnf}. In the second, a source creates an off-shell classical field of the new boson, and the experiment aims to detect the forces and torques caused by this classical field~\cite{Terrano:2015sna, Lee:2018vaq, Arvanitaki:2014dfa}.

The latter class of experiments are the focus of this paper. The interaction of the new boson with the standard model can either be through scalar and vector interactions or through pseudo-scalar (or pseudo-vector) interactions that couple to the spin of standard model particles such as nucleons and electrons. Direct scalar exchange between nucleons and electrons leads to long-range forces between these particles, which are heavily constrained by laboratory tests for new short distance forces~\cite{OHare:2020wah}. On the other hand, the best direct constraint on pseudo-scalar interactions arises from astrophysics~\cite{OHare:2020wah}. This leads to the possibility that if standard model particles have both scalar (``monopole'') and pseudo-scalar (via spin, or ``dipole'') interactions with this boson, then the product of these interactions could be better probed in the laboratory. The purpose of this paper is to propose a new search for this combination of ``monopole-dipole'' interactions.

To search for a new monopole-dipole interaction, one scheme is to place a source mass that produces the bosonic field through the monopole interaction. The bosonic field can be detected using a spin sensor, wherein the gradient of the bosonic field induces the spin to precess. Another scheme would be to use a spin-polarized sample which produces the bosonic field through the dipole interaction, and look for the bosonic field via the acceleration induced on a test body. Both of these are viable experimental approaches, and the ultimate sensitivity of each approach depends on technical considerations. 

Broadly, however, one can make the following observations. In an experiment where the produced bosonic field is detected with a spin, the sensitivity is set by the time for which the spin can freely precess before it relaxes. For nuclear spins, this time scale can be long (potentially as long as $\sim\! 1000\,\text{s}$), even when the spins are at high density such as in a liquid. Electron spin precession, however, is harder to probe through this method since the relaxation times of electrons are shorter. If one is interested in probing dipole interactions of the electron with the boson, it is thus enticing to look for the interaction by taking a spin-polarized sample of electrons and detecting the sourced field via the acceleration it causes on a test body. This concept has the additional advantage that a spin-polarized sample of electrons is simply a ferromagnet, which is readily obtainable, as opposed to the effort required to polarize nucleon spins (whether to source such a field or to detect it). To implement this concept, the acceleration induced by the field needs to be detected. Precision accelerometers work by permitting the acceleration to induce a position change on the test body and then using a suitable interferometric method to detect this length change. 
For terrestrial macroscopic test bodies, which must be attached to some support, the noise in that support (thermal or vibrational) imposes a limit on the time for which the test body can freely move, limiting the sensitivity. 

In this paper, we point out that this limitation of a macroscopic test body can be overcome by using freely falling atoms as the test body to detect the acceleration sourced by a suitably polarized spin sample. Such a system is naturally devoid of any support structure that would induce vibrational or thermal noise and thereby limit the free evolution time. Moreover, an atom has a well-defined response to relevant noise sources, such as magnetic field noise, permitting the amelioration of important systematic effects. Our primary interest in this paper will be monopole-dipole interactions where the new boson has monopole interactions with nucleons and dipole interactions with electrons. We will also comment on other possible combinations that can be probed with this technique.

The rest of this paper is organized as follows. We describe the main ideas of the experimental approach in Section~\ref{sec:approach} and outline the details of a possible experiment in Section~\ref{sec:setup}. We then discuss the expected sensitivity reach in Section~\ref{sec:sensitivity} and analyze leading systematic effects in Section~\ref{sec:systematics}.

\section{Experimental Approach}
\label{sec:approach}

\begin{figure}[t!]
\centering
\includegraphics[width=0.35\textwidth]{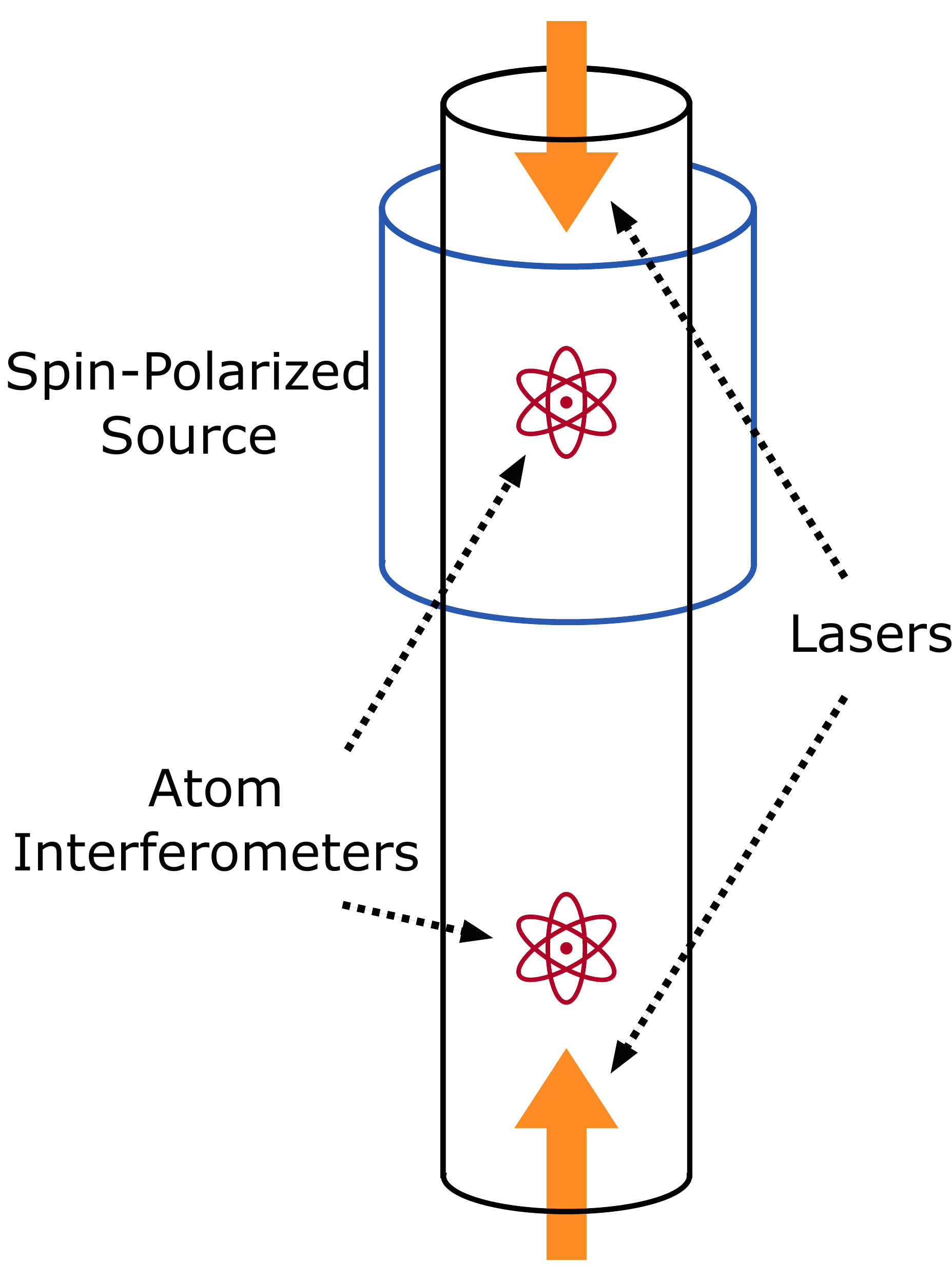}
\caption{Experimental concept. Two atom interferometers are placed in a gradiometer configuration, with one of the interferometers near a spin polarized source (such as a ferromagnet) and the other far from the source. The experimental signal is the differential phase shift between the two atom interferometers. This measurement is sensitive to any new force on the atoms while suppressing contributions from common sources of noise associated with the lasers.}
\label{Fig:Setup}
\end{figure}

The basic idea of the experiment is depicted in Fig.~\ref{Fig:Setup}. A spin-polarized source, such as a ferromagnet, produces the bosonic field. This source is placed near a magnetically shielded vacuum system in which two free-falling atom interferometers are operated in a gradiometer configuration, with one of the interferometers near the source and the other farther away. The phase shift of each interferometer is sensitive to the acceleration between the atoms and the optical phase reference (e.g. the laser delivery optics). In the gradiometer configuration, the differential phase shift between the two interferometers is sensitive to the relative acceleration between them. If the spin-polarized source exerts a new force on the atoms, it will cause a relative acceleration between the two interferometers since one of them is much closer to the source than the other.

With this setup, let us estimate the parameters of the new physics that the experiment can probe. We use the parameterization of~\cite{OHare:2020wah} to write the monopole-dipole potential induced by a new boson (such as an axion) between a nucleon and an electron that are at a distance $r$ from each other: 
\begin{equation}
V_{Ne}\left(r\right) = \frac{g_N g_{e}}{8 \pi m_e} \left( \frac{1}{r \lambda} + \frac{1}{r^2}\right) e^{-\frac{r}{\lambda}} \left(\hat{\sigma}\cdot\hat{r}\right).
\label{Eqn:Potential}
\end{equation}
Here $g_N$ is the scalar (or ``monopole'') coupling between a new boson and nuclei, $g_e$ is the spin-dependent (or ``dipole'') coupling between the boson and electrons, $m_e$ is the electron mass, and $\hat{\sigma}$ is a unit vector in the direction of the electron spin. The range of the new force is parameterized by $\lambda$. Since the force is carried by a new boson $\phi$ of mass $m_{\phi}$, $\lambda = h/m_{\phi} c$.

This experiment is anticipated to be sensitive to forces with $\lambda \gtrsim 10\,\text{cm}$. This expectation arises from multiple considerations. First, the atoms need to be placed inside a magnetically shielded vacuum region, away from the location of the spin-polarized source. The size of the vacuum chamber and shield will set a minimum distance from the atoms of several centimeters. Second, due to temperature, the transverse velocity of the atoms is $\sim\!1\,\text{cm/s}$. With a meter-scale spin source, the atoms will be near the source for $\sim\!1\,\text{s}$, implying that the transverse size of the atom clouds will be $\sim\!1\,\text{cm}$. The atom optics laser beam needs to be sufficiently homogeneous over this length scale, typically requiring a beam size of a few centimeters. Together, these requirements establish that the atom interferometers must be located at least $10\,\text{cm}$ away from the spin-polarized source, resulting in sensitivity to forces with a range $\lambda \gtrsim  10\,\text{cm}$. The experiment proposal considered in Sections \ref{sec:setup}--\ref{sec:systematics} assumes a minimum transverse distance of $20\,\text{cm}$ between the atoms and the spins.

The experiment will ideally be limited by the atom shot noise of the interferometers rather than by other systematic effects. The gradiometer configuration enables differential suppression of systematic effects that are common across the interferometers, such as noise associated with the laser or uniform background fields. However, backgrounds can also cause an acceleration gradient between the interferometers. The experiment should therefore have a way to establish that a signal (i.e., a differential phase shift between interferometers) is due to a new force from the spin-polarized source and not due to backgrounds. For example, this can be done by moving the source mass away from the interferometers to take null measurements without the bosonic field present, or by controlling the polarization of the source mass to change the direction of the new force. If the phase shift changes suitably under these actions, one can be confident that the spin-polarized source is the physical origin of the signal.

The differential measurement scheme described above suppresses any systematic effects uncorrelated with the spin-polarized source. However, it does not eliminate differential phase shifts arising from magnetic fields generated by the spin-polarized source. The need to suppress this systematic effect is a defining characteristic of the experiment. To tackle this problem, we will adopt two strategies: first, we will choose the geometry of the spin-polarized source and associated magnetic shields to suppress the induced magnetic field in the interferometer region; second, we will choose atom species and states with intrinsically small magnetic response.

We want to suppress the magnetic field in the interferometer region without suppressing the bosonic field from the source. This is possible because magnetic fields can be sourced by both spins and currents, whereas the new bosonic field is only produced by spins. Here we outline three possible strategies. First, the spin-polarized sample could be combined with an electromagnet in such a way as to cancel the net magnetic field in the interferometer region. This does not cancel the bosonic field in the interferometer region since the bosonic field is not sourced by the current in the electromagnet. We apply this strategy for the experimental setup proposed in Section~\ref{sec:setup}. Second, the interferometer region could be surrounded by a superconducting magnetic shield, with the spin-polarized source placed outside the shield. The magnetic fields from the source can then be cancelled by currents in the superconductor. Since these currents do not produce the bosonic field, this strategy could also suppress magnetic fields in the interferometer region without affecting the bosonic field. Third, the source of the bosonic field could be a magnet in which the magnetic field is produced by both spin and orbital angular momentum. Magnetic materials with different ratios of spin and orbital angular momentum can then be combined to suppress the magnetic field in the interferometer region while maintaining a net spin polarization, thus sourcing the bosonic field~\cite{Terrano:2015sna}.

Furthermore, we note that the magnetic field systematic effect can be independently characterized by using additional atom interferometers to measure the magnetic field in situ~\cite{Asenbaum:2020}. In such a configuration, one gradiometer would be operated in a magnetically insensitive state (e.g. the ${}^1\!S_0$ ground state of $^{88}$Sr) to measure the bosonic field, while another gradiometer would be operated in a magnetically sensitive state to measure the magnetic field. Such a system can decorrelate the phase shift caused by magnetic fields from the phase shift caused by the bosonic field, provided that the ratio of magnetic moments in the two gradiometers differs significantly from the ratio of their individual neutron/proton ratios. As we will show in Section~\ref{sec:systematics}, the magnetic field systematic effect is expected to be small enough that this co-magnetometry is likely to be unnecessary. Thus, we will focus on the sensitivity that could be achieved with a single gradiometer operated with a magnetically insensitive state.

In summary, we consider the setup in Fig.~\ref{Fig:Setup}, with a $\sim\!1\,\text{m}$ spin-polarized source mass and $\sim\!1\,\text{s}$ operation times for the free-falling atom interferometers. We will specifically investigate the use of $^{88}$Sr in this setup to mitigate the magnetic field systematic effect, in concert with a source mass that can electronically control its spin polarization to differentially suppress backgrounds.

\section{Setup}
\label{sec:setup}

\begin{figure}[t]
    \centering
    \includegraphics[width=0.35\textwidth]{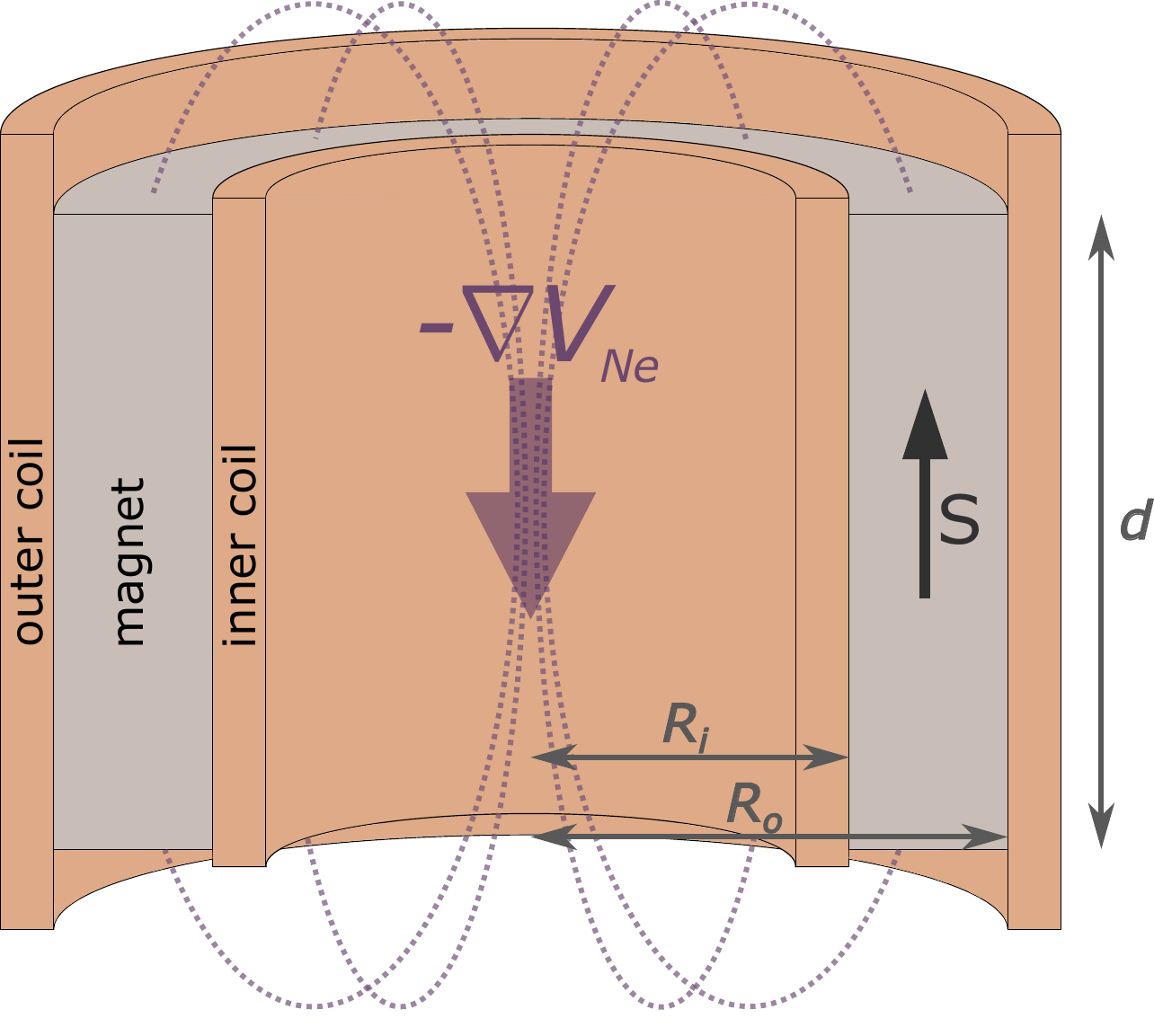}
    \caption{Cross-sectional diagram of a source mass designed to be placed around the atom interferometer. The source mass is an electromagnet consisting of a soft ferromagnet core of length $d$, with coils wound about its inner ($R_i$) and outer ($R_o$) radii. The outer coil is used to control the direction of the electron spins $S$ in the magnet. When the spins are axially polarized, the magnet generates a monopole-dipole force $-\nabla V_{Ne}$ along its center axis. The inner coil is used to locally suppress the magnetic fields generated by the source mass.}
    \label{fig:magnetschematic}
\end{figure}

In this section, we discuss a possible experiment setup following the principles described above. For example, the experimental parameters provided here would be compatible with a gradiometer apparatus currently under assembly at Stanford, which features two strontium interferometers vertically separated by $5\,\text{m}$~\cite{Garber:2024}. We provide sensitivity projections and systematics analysis based on this setup in Sections \ref{sec:sensitivity} and \ref{sec:systematics}.

A schematic of the spin-polarized source mass is shown in Fig.~\ref{fig:magnetschematic}. The bosonic field is sourced by the electron spins in the ferromagnetic core of an electromagnet. A strong magnet is desirable since the polarized spin density $n_s \approx \frac{B_\text{int}}{\mu_0 \mu_B}$ scales with the magnet's internal field $B_\text{int}$, where $\mu_0$ and $\mu_B$ are the vacuum permeability and the Bohr magneton. The core material should also have a high magnetic permeability for electromagnetically controlling the spins. Materials typically used in transformers can support $B_\text{int}\sim 10\,\text{kG}$ and $\mu \sim 10^3$--$10^4$. We consider a soft ferromagnet made of such a material in a hollow cylinder geometry, parameterized by its inner radius $R_i$, outer radius $R_o$, and height $d$. When the source mass is axially polarized, a gradient of the scalar potential $V_{Ne}$ (see Eq.~\eqref{Eqn:Potential}) is sourced in the center region of the magnet, where the atoms can detect the new force.

The source mass has two electromagnetic coils. First, an outer coil wound around $R_o$ is used to control the spin polarization. This enables differential measurements between multiple experiment configurations where the spins are reversed or randomized. Second, an inner coil wound around $R_i$ is used to cancel magnetic fields generated by the electromagnet. As discussed in Section~\ref{sec:approach}, this first order cancellation of the fields must be provided by free current and not by a high permeability ferromagnetic shield to avoid screening the bosonic field. Ideally, the inner coil should feature a spatially varying current density to suppress magnetic fields in the region where the interferometer is active. This can be achieved by dividing the inner coil into a discrete set of coils driven at different currents.

We consider a source mass magnetized to $B_\text{int}=5\,\text{kG}$, with geometric parameters $R_i=0.20\,\text{m}$, $R_o=0.23\,\text{m}$, and $d=0.50\,\text{m}$. The magnetization and geometry are selected based on practical considerations for the magnetic and gravitational systematics in the experiment. When magnetized, the source mass generates a $\sim\!220\,\text{G}$ field at the center of the hollow cylinder, which can be compensated with modest currents on the inner coil. Assuming a mass density of $8\,\text{g/cm}^3$ (typical for iron-based magnetic alloys) the magnet weighs $\sim\!160\,\text{kg}$. A stronger and larger source mass is strictly better for sourcing the bosonic field, but becomes impractical beyond a certain scale for the setup being considered.

\begin{figure}[t]
    \centering
    \includegraphics[width=0.5\textwidth]{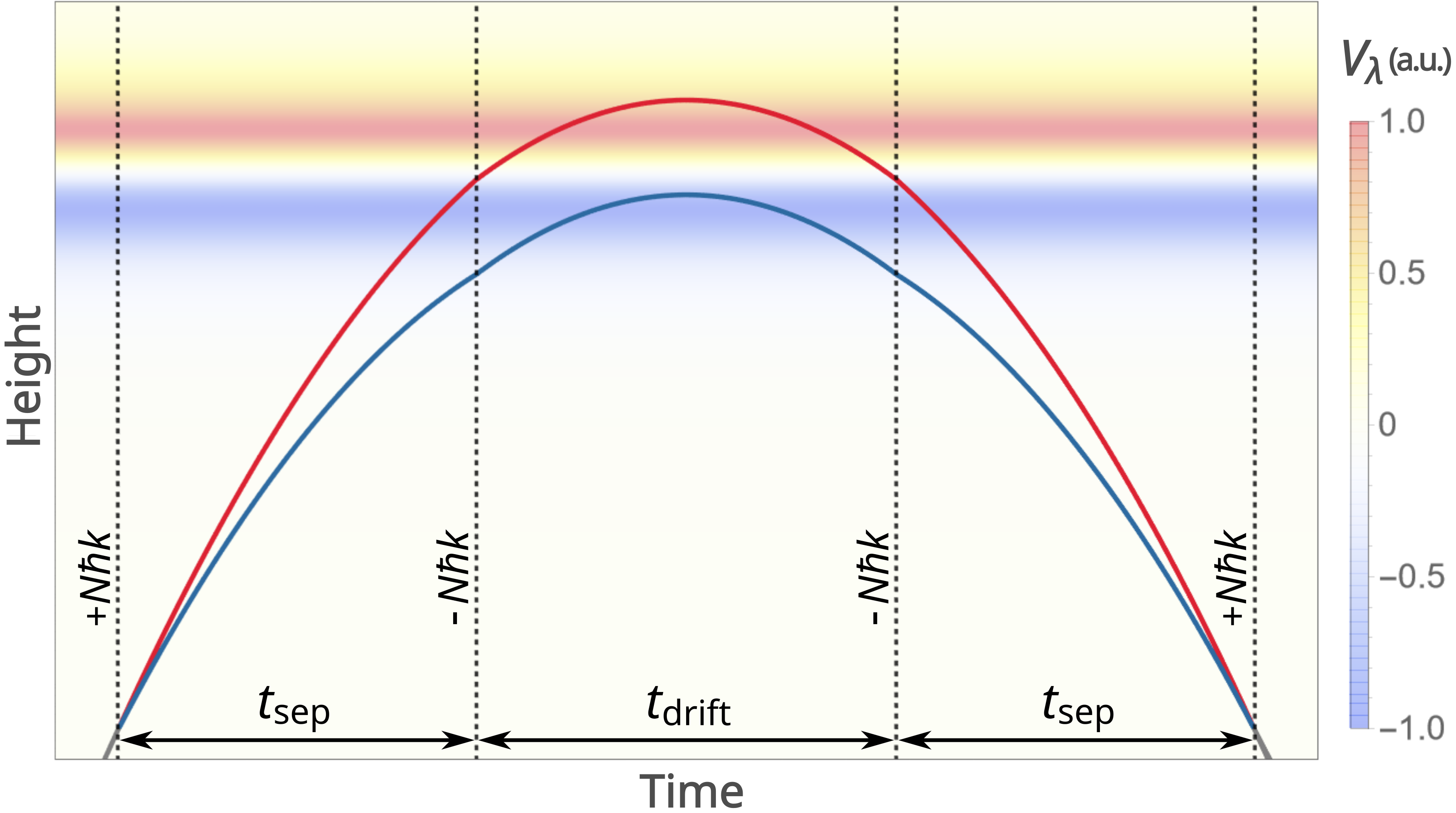}
    \caption{Example spacetime diagram of the proposed atom interferometer experiment. The reference interferometer for the gradiometer configuration is not shown. The red and blue lines show the upper and lower arm trajectories of the interferometer. Dashed lines show locations where laser pulses are applied. The interferometer is opened with a beamsplitter pulse, followed by a $+N \hbar k$ LMT pulse sequence. After a time $t_\text{sep}$, a $-N \hbar k$ sequence is applied to reduce the differential velocity between the arms. After an additional time $t_\text{drift}$, another $-N \hbar k$ sequence and $t_\text{sep}$ reverse the arm separation, followed by a $+N \hbar k$ sequence and final beamsplitter pulse to close the interferometer. The overlaid heatmap corresponds to the potential seen by the atoms due to a spin-sourced bosonic field generated by the magnet described previously (see Eq.~\eqref{Eqn:CylinderPotential}).}
    \label{fig:spacetimediagram1}
\end{figure}

Fig.~\ref{fig:spacetimediagram1} shows a spacetime diagram of an atom interferometer sequence probing the region of the source mass. First, the atoms are launched upward and split into two trajectories via a beamsplitter pulse. These interferometer arms are then addressed with a large momentum transfer (LMT) sequence~\cite{Rudolph:2020} consisting of $N$ photon momentum kicks, followed by a wait time $t_{\text{sep}}$. The parameters $N$ and $t_{\text{sep}}$ are selected to spatially separate the interferometer arms by $\sim\!d$. Next, a second LMT sequence reduces the momentum separation and the arms spend a time $t_{\text{drift}}$ following near-parallel trajectories in the vicinity of the source mass. Lastly, the inverse pulse sequences are applied to close the interferometer. Qualitatively, this sequence can be interpreted as the interferometer arms probing opposite ends of the source mass, where the potential difference due to the bosonic field is greatest. In the next section, we show that the sequence parameters can be optimized for the phase shift caused by this potential difference. We find the optimized parameters $N=170$, $t_{\text{sep}}=0.57\,\text{s}$, $t_{\text{drift}}=0.67\,\text{s}$, with the source mass positioned at a height $3.85\,\text{m}$ above the initial position of the atoms. The arm separation during the drift time is $0.65\,\text{m}$. Similar atom interferometer arm separations have been demonstrated in~\cite{Kovachy:2015}.

In order to suppress the magnetic systematic effects associated with the source mass, the atom interferometer should operate on magnetically insensitive states of a bosonic isotope. We propose a $^{88}$Sr interferometer driven on the ${}^1\!S_0-{}^3\!P_1(m=0)$ transition, for which LMT sequences have been demonstrated up to $N=400$~\cite{Rudolph:2020, Wilkason:2022}. The interferometer arms can be shelved in the ${}^1\!S_0$ ground state during the free propagation time of the sequence, both to circumvent spontaneous emission losses and to mitigate ${}^3\!P_1(m=0)$ excited state coupling to magnetic fields~\cite{Rudolph:2020}.

With the source mass and interferometer described above, one can take the approach outlined in Section~\ref{sec:approach}, where the signal is distinguished from systematic effects both by gradiometric measurements and by differential measurements with the spins reversed. We note that one could also take advantage of the gradiometer by positioning source masses of opposite polarity at each of the two interferometer locations, resulting in a factor of two enhancement of the signal.

\section{Sensitivity}
\label{sec:sensitivity}

In this section, we estimate the sensitivity of the proposed experiment to the fundamental parameters $g_{N} g_{e}$ (see Eq.~\eqref{Eqn:Potential}) as a function of the range $\lambda$ of the new force.

In principle, when the experimental apparatus is fully described, the bosonic field in the interferometer region can be completely calculated. This would require specification of both the spin-polarized source and any magnetic shielding, since the net field in the interferometer region also depends on the spins in the magnetic shield. Such a calculation is beyond the scope of this paper, and we will instead make a simplified estimate to illustrate the potential reach of the experiment.

For this estimate, we use the source mass geometry described in Section~\ref{sec:setup} and calculate the potential given by Eq.~\eqref{Eqn:Potential} assuming a uniform, axial spin polarization. We obtain the monopole-dipole potential for an atom along the axis of the cylinder: 
\begin{equation} \label{Eqn:CylinderPotential}
    V_\lambda(z) = \frac{g_{N} g_e n_s A \lambda} {4 \, m_e}
    \left(
    e^{-\frac{\sqrt{R_i^2+(z-d/2)^2}}{\lambda}}
    -
    e^{-\frac{\sqrt{R_i^2+(z+d/2)^2}}{\lambda}}
    +
    e^{-\frac{\sqrt{R_o^2+(z+d/2)^2}}{\lambda}}
    -
    e^{-\frac{\sqrt{R_o^2+(z-d/2)^2}}{\lambda}}
    \right).
\end{equation}
\noindent Here $n_s$ is the number density of spins in the source mass, $A$ is the number of nucleons in the atom, and $z$ is the position along the axis of the cylinder. We then calculate the induced phase shift in the atom interferometer by integrating the monopole-dipole potential along the interferometer arm trajectories $z_1(t)$ and $z_2(t)$~\cite{CCT:1994}: 
\begin{equation}  \label{Eqn:signalphase}
    \Delta \phi_{\lambda}
    = \frac{1}{\hbar} \int \Big(V_\lambda(z_1(t)) - V_\lambda(z_2(t))\Big)\, dt.
\end{equation}
The phase sensitivity in this configuration saturates when the interferometer arm separation $z_1 - z_2$ becomes comparable to the height of the source $d$. We assume that the second interferometer in the gradiometer is located far enough below the source that the effect of the monopole-dipole potential on its phase is negligible. 

\begin{figure}[t]
\centering
\includegraphics[width=0.5\textwidth]
{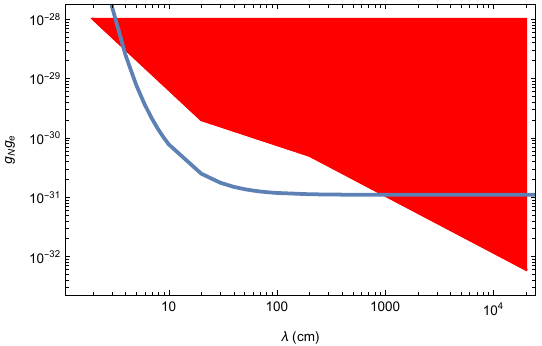}
\caption{Sensitivity of the monopole-dipole interaction strength, characterized by the coupling parameters $g_N g_e$, as a function of the interaction range $\lambda$. The red shaded region indicates the parameter space rejected by previous laboratory experiments~\cite{OHare:2020wah}. The blue curve shows the projected sensitivity of the experiment setup described in Section~\ref{sec:setup}. The experiment has the potential to surpass existing bounds in the range $\lambda\sim 10$--$10^3\,\text{cm}$.}
\label{Fig:sensitivity}
\end{figure}

Fig.~\ref{Fig:sensitivity} shows the projected sensitivity of the proposed experiment to the monopole-dipole interaction as a function of the range $\lambda$, assuming a phase resolution of $10^{-3}\,\text{rad}/\sqrt{\text{Hz}}$ and a one-year measurement campaign. 
 Existing bounds on the product $g_N g_e$ from laboratory experiments~\cite{OHare:2020wah} are also shown. Laboratory experiments currently set the strongest constraints on $g_N$. There are astrophysical limits on $g_e$ that are stronger than the laboratory limits, which would constrain the product $g_N g_e$ to $\mathcal{O}\left(10^{-35} - 10^{-34}\right)$ over the length scales of interest in this experiment. However, these astrophysical limits are model-dependent~\cite{DeRocco:2020xdt}, and a model-independent constraint requires direct experimental tests in known environments.

\section{Systematics}
\label{sec:systematics}

This experiment involves positioning a ferromagnetic source mass near an atom interferometer. As such, an important systematic error is expected from shifts in the atomic energy levels due to magnetic fields. To reduce this error we use the ${}^1\!S_0$ ground and ${}^3\!P_1(m=0)$ excited states of $^{88}$Sr, which do not have a first-order Zeeman shift. The leading order energy shift for these states are then quadratic in the field amplitude, $\alpha_i |\Vec{B}_\text{tot}|^2$, where $\alpha_i$ is the quadratic magnetic susceptibility of atomic state $i$. To estimate the resulting phase shifts, we first calculate the magnetic fields present at the interferometer. The polarized spins of the source mass described in Section~\ref{sec:setup} can generate a field $B_\text{mag}\sim 220\,\text{G}$ along its center axis. With appropriate current tuning along the inner coil, we assume this field can be mitigated to the percent level in a $1\,\text{m}$ interferometer region of interest. We also assume that the atom interferometer region is surrounded by a magnetic shield. For example, the Stanford apparatus is equipped with a magnetic shield designed to suppress the Earth field to sub-mG levels~\cite{Garber:2024, Abe:2021}. We consider a combined magnetic field suppression factor of $10^5$ from the coils and shield, leaving a remaining residual field $B_\text{res}\sim 2.2\,\text{mG}$ from the source mass. In addition, a transverse bias field $B_\text{bias} \sim 1\,\text{G}$~\cite{Abe:2021} is needed to set the quantization axis for the interferometer. The total magnetic field at the interferometer is thus the vector sum $\Vec{B}_\text{tot}=\Vec{B}_\text{res}+\Vec{B}_\text{bias}$.

Based on this field, we estimate the worst case magnetic phase shift where only one interferometer arm experiences the maximal residual field. The resulting set of phase shift terms introduced by the source mass are summarized in Table \ref{tab:phaseshiftsummary}. The terms are either quadratic in the residual field $|\Vec{B}_\text{res}|^2$ or product terms of the residual and bias fields $\Vec{B}_\text{bias} \cdot \Vec{B}_\text{res}$, and are separated into their phase contributions from the ground and excited states. For the phase accumulated in the ground state, we estimate a magnetic field susceptibility of $\alpha_{g} \approx 2 \pi \times 6\,\text{mHz/G}^2$~\cite{DelAguila:2018,*DelAguila:2019Corrigendum} and a time $t_g \approx t_\text{drift}$ spent in the presence of the residual magnetic field. Similarly for the phase accumulated in the excited state, we estimate a susceptibility $\alpha_{e} \approx 2 \pi \times 60\,\text{mHz/G}^2$~\cite{Ferrari:2003} and a time $t_e \approx N t_\pi$, where $t_\pi=100\,\text{ns}$ is the single pulse duration and the total duration comes from the two pulse sequences applied near the magnet. We also take into account that the $\Vec{B}_\text{bias} \cdot \Vec{B}_\text{res}$ product depends on the angle between a primarily axial $\Vec{B}_\text{res}$ and a transverse $\Vec{B}_\text{bias}$. In practice, the transverse component of the residual field depends on details of the shielding setup. For this analysis, we estimate the transverse component of $\Vec{B}_\text{res}$ from the average magnetic field gradient along the axis of a uniformly magnetized source mass. The product terms are then reduced by a factor of $0.8\,\frac{x}{d}\sim 10^{-3}$, where $x\sim 1\,\text{mm}$ is the atom cloud size.

Notably, the magnetic field phase shift terms in Table~\ref{tab:phaseshiftsummary} can be decorrelated from the signal phase shift $\Delta \phi_\lambda$ by taking a suitable set of differential measurements. For the $|\Vec{B}_\text{res}|^2$ terms, reversing the source mass spin polarization changes the sign of $\Delta \phi_\lambda$ but not of the magnetic field phase shift. For the $\Vec{B}_\text{bias} \cdot \Vec{B}_\text{res}$ terms, reversing the direction of the bias field changes the sign of the magnetic field phase shift but not of $\Delta \phi_\lambda$. The magnetic field phase shifts can therefore be differentially suppressed by taking measurements that toggle the spin direction and the bias field direction. We assume percent-level suppression from these strategies based on the magnetic field control we have at the atoms, limited by current noise on the bias and source mass coils. The magnetic field phase shift values, with and without differential suppression, are summarized in Table~\ref{tab:phaseshiftsummary}. The suppressed phase shift terms are projected to be below the final sensitivity and should not limit the experiment.

\begingroup
\setlength{\tabcolsep}{6pt} 
\renewcommand{\arraystretch}{1.5} 
\begin{table}
\begin{center}
\begin{tabular}{c c c c}
\hline
&
Phase Shift & 
Value (Raw) &
Value (Suppressed) \\
\hline \hline 
Signal
&
$\Delta \phi_{\lambda}$
&
$2 \times 10^{-7}$
&
$2 \times 10^{-7}$ \\
\hline
Magnetic field
&
$2 \alpha_{g} t_{g} \Vec{B}_\text{bias} \cdot \Vec{B}_\text{res}$
&
$2 \times 10^{-7}$
&
$3 \times 10^{-9}$
\\
&
$\alpha_{g} t_{g} |\Vec{B}_\text{res}| ^2$
&
$1 \times 10^{-7}$
&
$2 \times 10^{-9}$
\\
&
$2 \alpha_{e} t_{e} \Vec{B}_\text{bias} \cdot \Vec{B}_\text{res}$
&
$4 \times 10^{-11}$
&
$9 \times 10^{-13}$
\\
&
$\alpha_{e} t_{e} |\Vec{B}_\text{res}| ^2$
&
$3 \times 10^{-11}$
&
$6 \times 10^{-13}$
\\
\hline
Gravity &
$\Delta \phi _\text{gg}$
&
$1 \times 10^{-7}$
&
$1 \times 10^{-11}$
\\
\hline  \\
\end{tabular}
\end{center}

\caption{Estimated values for the projected signal and major systematic phase shifts for the proposed experiment after a one-year measurement campaign. $\Delta \phi_{\lambda}$ and $\Delta \phi _\text{gg}$ are numerically calculated from Eq.~\eqref{Eqn:signalphase} and Eq.~\eqref{Eqn:PhiGravAxial}. The 
magnetic phase shifts are discussed in Section~\ref{sec:systematics} and include differential and geometric suppression factors. \label{tab:phaseshiftsummary}}
\end{table}
\endgroup

We also consider the phase shift due to the gravitational potential $V_\text{grav}$ of the $160\,\text{kg}$ source mass. This effect, like other backgrounds, is suppressed by taking differential measurements with reversed spin orientations. However, shot-to-shot variations in the relative position $\Delta z$ between the atoms and source mass can cause fluctuations in the phase shift of the form
\begin{equation}
    \Delta \phi _{\text{gg}} = 
    \frac{1}{\hbar}
    \left(\int \Big( V_{\text{grav}}(z_1(t) + \Delta z) -
    V_{\text{grav}}(z_2(t) + \Delta z) \Big) \,dt
    - \int \Big(V_{\text{grav}}(z_1(t)) - V_{\text{grav}}(z_2(t)) \Big)\, dt \right).
    \label{Eqn:PhiGravAxial}
\end{equation}
\noindent
The strength of this effect depends on $\Delta z$ and on the local gravity gradients from the source mass. We assume $\Delta z=10\, \mu \text{m}$ arising from limitations in the mechanical stability of a typical lab environment, for example jitter in the source mass mounting or initial position uncertainty of the atoms. Evaluating Eq.~\eqref{Eqn:PhiGravAxial} along the interferometer trajectories we find an uncertainty $\Delta \phi _{\text{gg}}=1 \times 10^{-7}$ after one year of measurement time, assuming that the position uncertainty can be averaged down over this time scale.  This uncertainty is similar in size to the shot noise and potentially limits the experimental sensitivity (see Table \ref{tab:phaseshiftsummary}). We note that the strongest $V_\lambda$ and the steepest gravity gradient both occur near the ends of the source mass; thus, trajectories optimized for $\Delta \phi _{\lambda}$ are also maximally susceptible to $\Delta \phi _{\text{gg}}$.

There are several strategies that could mitigate this effect and relax the requirements. First, one could take null measurements with an unpolarized source mass while monitoring the source mass position (e.g. with a laser interferometer) and the atom cloud position (e.g. imaging the atoms). $\Delta \phi _{\text{gg}}$ could then be decorrelated by characterizing the atom interferometer phase as a function of the monitored heights. In this case $\Delta z$ is set by the measurement uncertainty of the monitoring devices. Second, one could spatially separate the gravity gradient from $V_\lambda$ by attaching non-magnetic trim masses to the source mass ends. These trim masses would flatten the gravitational field profile over the active region of the interferometer, assuming they can be rigidly fixed to the magnetic mass within a $10\, \mu \text{m}$ tolerance.

Alternatively, the gravitational systematic can be addressed with frequency shift gravity gradient (FSGG) compensation~\cite{DAmico:2017,Overstreet:2018}. In this method, laser pulses are frequency-shifted to imprint a phase that cancels the interferometer phase proportional to the initial position and velocity of the atoms. For the source mass and interferometer considered here, we find that $\Delta \phi _{\text{gg}}$ is proportional to $\Delta z$ over a range of several millimeters. That is,
\begin{equation}
    \Delta \phi _{\text{gg}}
    \approx \frac{1}{\hbar} \Delta z \!\int\!\left( \frac{\partial V_{\text{grav}}(z_1(t))}{\partial z} - \frac{\partial V_{\text{grav}}(z_2(t))}{\partial z} \right)\, dt.
\label{Eqn:LinearPhiGravAxial}
\end{equation}
We also find that compensating for this phase shift requires a frequency shift of $\sim\!2\,\text{GHz}$, which can be done with Bragg or Raman pulses. Therefore, FSGG compensation could be implemented for this experiment in a hybrid interferometer, using Bragg transitions on ${}^1\!S_0-{}^1\!P_1$~\cite{Mazzoni:2015} for beamsplitter and mirror pulses and using single-photon transitions on ${}^1\!S_0-{}^3\!P_1$ for LMT pulses~\cite{Rudolph:2020,Wilkason:2022}. The compensation Bragg pulses can be calibrated by scanning over a range of interferometer starting positions and measuring the gravity gradient phase shift of an unpolarized source mass. With shot-noise-limited operation, several hours of calibration could then suppress $\Delta \phi_{\text{gg}}$ by a factor of $10^4$. In Table \ref{tab:phaseshiftsummary}, we summarize the suppressed and unsuppressed gravitational effect assuming FSGG compensation to this level.

\begin{figure}[t]
    \centering
    \includegraphics[width=0.5\textwidth]{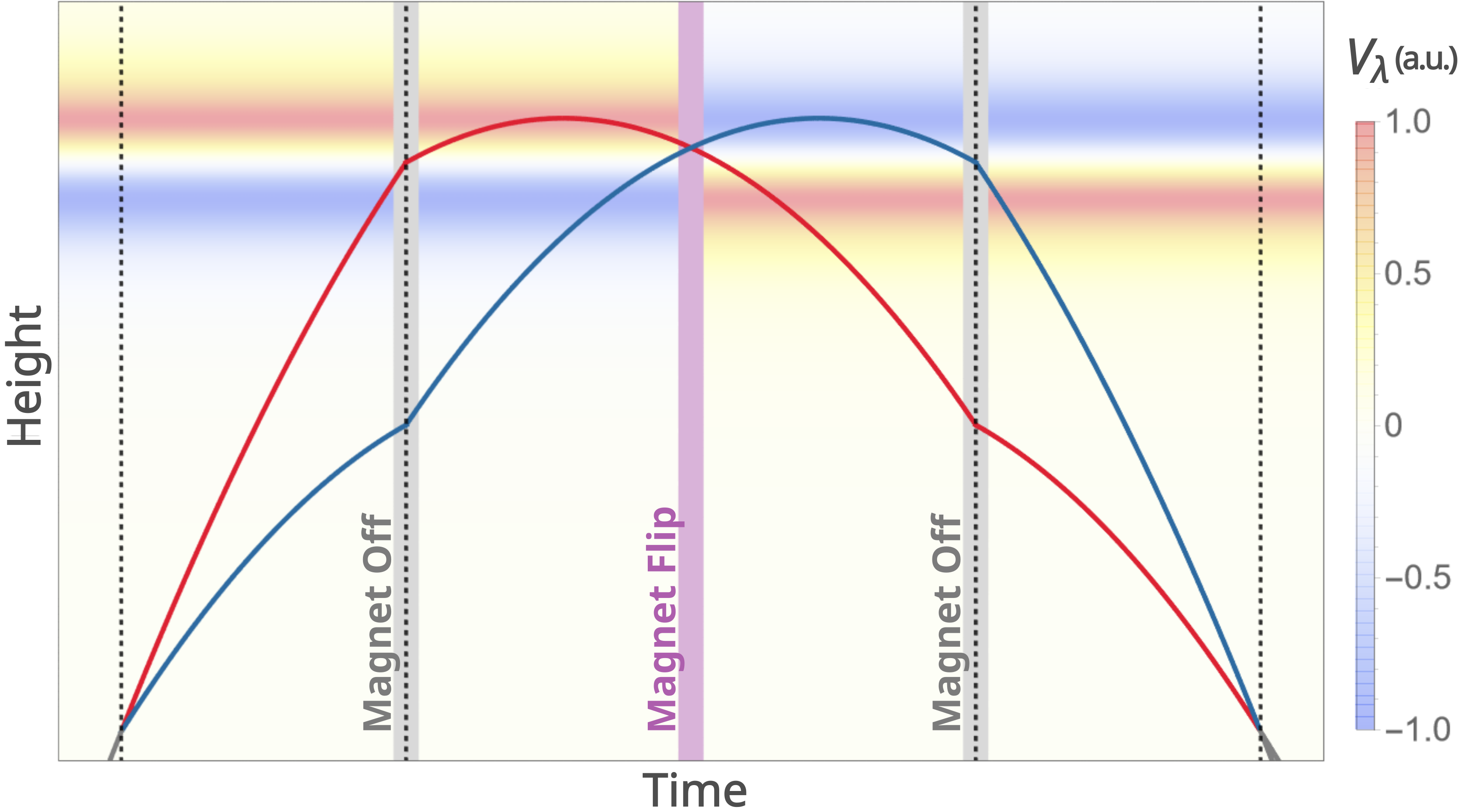}
    \caption{Example spacetime diagram of a two-diamond atom interferometer sequence for suppressing systematic effects. The red and blue lines show the two interferometer arm trajectories, and the dashed lines show the locations of the laser pulse sequences. Gray regions indicate when the magnet polarization is switched off to avoid excited state coupling with magnetic fields. The purple region indicates when the magnet polarization is being reversed. By changing the sign of the bosonic field potential when the interferometer arms cross, the signal phase is accrued coherently while low frequency background noise is suppressed.}
\label{fig:spacetimediagram2}
\end{figure}

Finally, we note that systematic effects in the experiment could be further suppressed with a source mass designed to control its spins during a single interferometry sequence. Such a source mass enables a double-loop sequence (as shown in Fig.~\ref{fig:spacetimediagram2}) where the spin polarizations are reversed and the interferometer arms are swapped synchronously. The symmetry of multi-loop interferometer trajectories is useful for suppressing low frequency backgrounds~\cite{Dubetsky:2006,Graham:2016}. Here, the double-loop sequence would suppress both the magnetic and gravity gradient phase shifts in Table~\ref{tab:phaseshiftsummary} within each shot, potentially removing the need for the other suppression strategies discussed above. Additionally, fast control of the spin polarity could be used to demagnetize the source mass during the interferometry pulses. This would avoid excited state coupling to leakage magnetic fields, easing the requirement of having a magnetically insensitive interferometer. Fig.~\ref{fig:spacetimediagram2} shows a sequence implementing these strategies. The source mass in this case must be capable of reversing its polarity at millisecond timescales without generating significant transient magnetic or gravitational fields.

\section{Conclusions}
\label{sec:conclusions}
The atom interferometry experiment considered in this paper has the potential to surpass existing laboratory probes of monopole-dipole forces between nucleons and electrons in the range $\lambda \sim 10$--$10^3\,\text{cm}$ by an order of magnitude. This method is complementary to experimental efforts searching for monopole-dipole forces between nucleons~\cite{Arvanitaki:2014dfa, Lee:2018vaq}. One could consider a protocol similar to our proposed experiment to look for monopole-dipole forces between nuclei by replacing the ferromagnet with a sample of polarized nuclear spins, with the field sourced by the spins detected with an atom gradiometer. In that case, the expected sensitivity of the experiment ($\sim\!10^{-28}$ on the product of the dimensionless nucleon monopole-dipole coupling, as parameterized by~\cite{OHare:2020wah}) is comparable to current laboratory limits on such forces. Given the complexities involved in achieving $\mathcal{O}\left(1\right)$ nuclear spin polarization, our method could be competitive if atom interferometer sensitivities significantly improve. In addition to these terrestrial detection possibilities, it would also be interesting to investigate potential satellite-based experiments. In orbit, the free fall times could be considerably longer, likely resulting in enhanced sensitivity. Given the ongoing interest in realizing quantum sensing platforms in space (such as~\cite{elliott2018nasa}), this possibility deserves further investigation.

\section*{Acknowledgments}
This work was supported by the U.S.~Department of Energy~(DOE), Office of Science, National Quantum Information Science Research Centers, Superconducting Quantum Materials and Systems Center~(SQMS) under Contract No.~DE-AC02-07CH11359. D.E.K.~and S.R.~are supported in part by the U.S.~National Science Foundation~(NSF) under Grant No.~PHY-2412361.
S.R.~is also supported by the Simons Investigator Grant No.~827042, and by the~DOE under a QuantISED grant for MAGIS. 
D.E.K.~is also supported by the Simons Investigator Grant No.~144924. M.A and J.M.H. are supported by the Gordon and Betty Moore Foundation Grant GBMF7945.

\bibliography{references}
\end{document}